\documentstyle[prd,aps,floats]{revtex}

\def\be{\begin{equation}}
\def\ee{\end{equation}}
\def\bea{\begin{eqnarray}}
\def\eea{\end{eqnarray}}

\def\npb#1#2#3{{\it Nucl.\ Phys.} {\bf B#1} (#2) #3}
\def\plb#1#2#3{{\it Phys.\ Lett.} {\bf B#1} (#2) #3}
\def\prl#1#2#3{{\it Phys.\ Rev.\ Lett.} {\bf #1} (#2) #3}

\def\rmp#1#2#3{{\it Rev.\ Mod.\ Phys.} {\bf #1} (#2) #3}
\def\mpla#1#2#3{{\it Mod.\ Phys.\ Lett.} {\bf A#1} (#2) #3}

\def\jhep#1#2#3{{\it JHEP\/} {\bf #1} (#2) #3}
\def\atmp#1#2#3{{\it Adv.\ Theor.\ Math.\ Phys.} {\bf #1} (#2) #3}

\def\p{\partial}
\def\tilde#1{\widetilde{#1}}

\def\CD{{\cal D}}                   \def\CG{{\cal G}}
                   
                   \def\CL{{\cal L}}
                   \def\CN{{\cal N}}

\def\C{{\bf C}}                     
\def\R{{\bf R}}                     
                     \def\Z{{\bf Z}}
\def\bx{{\bf x}}
\def\bk{{\bf k}}
\def\bp{{\bf p}}
\def\K{{\rm K}}

\def\KR{{\rm KR}}

\begin{document}

\input epsf
\renewcommand{\topfraction}{0.99}
\renewcommand{\bottomfraction}{0.99}
\twocolumn[\hsize\textwidth\columnwidth\hsize\csname
@twocolumnfalse\endcsname

\title{Stability of Fermi surfaces and K-theory}
\author{Petr Ho\v{r}ava}
\address{
Berkeley Center for Theoretical Physics and Department of Physics,\\ 
University of California,
Berkeley, California 94720-7300\\ 
and\\
Theoretical Physics Group, Lawrence Berkeley National Laboratory,\\ 
Berkeley, California 94720-8162}
\date{\today}
\maketitle
\begin{abstract}
Nonrelativistic Fermi liquids in $d+1$ dimensions exhibit generalized Fermi 
surfaces: $(d-p)$-dimensional submanifolds in the $(\bk,\omega)$-space 
supporting gapless excitations.  We show that the universality classes of 
stable Fermi surfaces are classified by K-theory, with the pattern of 
stability determined by Bott periodicity.  The Atiyah-Bott-Shapiro 
construction implies that the low-energy modes near a Fermi surface exhibit 
relativistic invariance in the transverse $p+1$ dimensions.  This suggests 
an intriguing parallel between norelativistic Fermi liquids and D-branes of  
string theory.
\end{abstract}


\vskip2pc]

The main focus of this paper is on non-relativistic condensed matter systems.  
However, its primary motivation comes from recent findings in superstring 
theory, in particular the physics of D-branes.  

D-branes are non-perturbative, inherently stringy solitonic defects in Type I 
and Type II superstring theory, defined as hypersurfaces in space-time on 
which strings can end \cite{dbranes}.  They naturally couple \cite{joep} to 
higher-form gauge fields (Ramond-Ramond fields) in superstring theory, in a 
manner analogous to how charged particles couple to the electromagnetic 
potential.  

Given a space-time background $Y$ that solves string theory, it is natural to 
attempt a classification of D-brane charges on $Y$.  The Ramond-Ramond (RR) 
fields that D-branes couple to have traditionally been viewed as differential 
forms on $Y$.  At low energies, the theory is described by a supergravity 
theory, whose action is schematically given by \cite{joebook}
\be
\CL\sim\int_Y d^{10}x \sqrt{g}\left(R-\sum_p\frac{1}{p!}dC_p\wedge\star dC_p+
\ldots\right).
\label{eesugralag}
\ee
The allowed degrees $p$ of the RR forms $C_p$ depend on the Type of the 
superstring theory considered. Thus, it may seem natural to expect that 
charges of D-branes take values in the de~Rham cohomology $H^\ast(Y,\R)$, or 
perhaps its generalization over $\Z$ when the Dirac charge quantization is 
taken into account.  

This expectation is now known to be incorrect \cite{mm,ewk,phk}.  D-branes 
are stringy objects, and not just submanifolds in space-time, and open string 
excitations give the D-brane world-volume $\Sigma$ extra structure.  In 
particular, $\CN$ coincident D-branes wrapping $\Sigma$ are characterized by 
a $U(\CN)$ gauge bundle on $\Sigma$.  When these facts are properly taken 
into account, it turns out that D-brane charges on a given manifold $Y$ take 
values in K-theory groups of $Y$.  K-theory (see e.g.~\cite{karoubi,spingeo} 
for some background) is an exotic cohomology theory, built from equivalence 
classes of $U(\CN)$ bundles on $Y$, and closely related to but distinct from 
the conventional cohomology $H^\ast(Y,\Z)$.

This implies that the RR fields are also objects in K-theory, and not 
differential forms \cite{mw}, making the low-energy description of string 
theory on $Y$ in terms of (\ref{eesugralag}) questionable.  The operations 
needed in (\ref{eesugralag}) -- the exterior derivative $d$, the wedge product 
$\wedge$, and the Hodge star $\star$ -- are well-defined on differential 
forms, but once the $C_p$'s are reinterpreted as K-theory objects, it is not 
clear how to even define (\ref{eesugralag}).  This crisis of the Lagrangian 
formulation of low-energy string theory is further supported by the discovery 
\cite{dmw} of apparently non-Lagrangian phases in the partition functions 
of various string and M-theory vacua.  

Perhaps this means that the Lagrangian framework currently available is 
insufficient for RR fields but its suitable generalization awaits to be 
discovered.  (Important steps in this direction have been taken 
\cite{freed}.)  Alternatively, the theory may require a non-Lagrangian 
formulation.  (This may already be suggested by the presence of a self-dual 
RR field strength in Type IIB theory).   Before we settle on either of these 
two alternatives, however, we should consider a third possibility.  The subtle 
K-theory features of string theory could be an emergent phenomenon, with 
D-branes and RR fields emerging as composites of some more elementary degrees 
of freedom that admit a conventional Lagrangian description.  

One purpose of this paper is to demonstrate that in another area of physics -- 
namely, in norelativistic condensed matter -- a close analog of this third 
possibility is realized:  Objects classified by K-theory do emerge as derived 
objects in systems whose microscopic degrees of freedom conform to a very 
conventional Lagrangian path-integral description.  In the process, we will 
learn potentially valuable lessons for condensed matter systems, and will 
discover an intriguing parallel between Fermi liquids and D-branes of string 
theory.  

In the rest of this paper, we will study non-relativistic Fermi liquids in 
$d+1$ space-time dimensions.  Usually, a restriction to $d\leq 3$ quickly 
follows, but we will keep $d$ arbitrary.  This will make some of the patterns 
more transparent; moreover, string theory is itself naturally formulated in 
space-times of higher dimensionality. 

We are interested in the low-energy dynamics of long-wavelength excitations 
in a Fermi liquid, described microscopically by a complex fermion 
$\psi^{i\sigma}(\bx,t)$.  $\sigma$ is the spinor index of the $SO(d)$ rotation 
group, and $i$ is an internal index of an $n$-dimensional representation 
of some compact symmetry $G$.  For simplicity, we consider $i$ the fundamental 
of $SU(n)$, the generalization to any compact $G$ being straightforward.  
Since the irreducible spinor representation of $SO(d)$ is 
$2^{[d/2]}$-dimensional (with $[k]$ denoting the integral part of $k$), 
$\psi$ has $N\equiv 2^{[d/2]}\cdot n$ complex components.  Real fermions will 
be considered in the closing part of the paper.  

The theory is microscopically described by the following non-relativistic 
Lagrangian, 
\bea
S&=&-\int dt\;d^d\bx\left(i\psi^\dagger_{i\sigma}\p_t\psi^{i\sigma}+
\frac{1}{2m}\psi^\dagger_{i\sigma}\Delta\psi^{i\sigma}
-\mu\;\psi^\dagger_{i\sigma}\psi^{i\sigma}\right)\nonumber\\
&&\qquad\qquad\qquad\qquad\qquad{}+\ldots
\nonumber 
\eea
Here $\Delta$ is the Laplace operator in $d$ dimensions, $\mu$ is the 
chemical potential, and ``$\ldots$'' denotes all interactions, such as those 
with an order parameter, four-Fermi interactions, impurities, etc.  

We wish to identify possible universality classes of the RG behavior and the 
corresponding RG fixed points.  The essence of the Landau theory of Fermi 
liquids \cite{shankar,joef} is in the observation that the low-energy 
excitations of the system populate a narrow shell near the Fermi surface, 
suggesting a coarse-graining procedure whereby momenta are scaled towards the 
Fermi surface.  Free fermions represent the simplest fixed point of this 
scaling.  The Fermi surface supports gapless coarse-grained fermionic 
excitations $\chi$, with the characteristic linear dispersion relation 
$\omega\sim k+\ldots$ at low energies.  

Once a free-field fixed point has been identified, a full RG analysis 
reveals possible instabilities of the system, due to (marginally) relevant 
interactions.  The degeneracy of the gapless modes $\chi$ can be lifted 
completely, and the system can develop a gap (such as in the case of 
superconductivity, or in the B-phase of ${}^3$He).  Alternatively, 
the gapless excitations can be only partially lifted, leaving a submanifold 
of some lower dimension $(d-p)$ in the $(\bk,\omega)$-space where gapless 
fermionic excitations are supported.  Since such submanifolds with gapless 
excitations generalize the concept of the Fermi surface, we will refer to them 
as ``generalized Fermi surfaces'' (and drop the word ``generalized'' from now 
on).  

Perhaps the standard example of a partially lifted Fermi surface is the A 
phase of ${}^3$He in $d=3$, where the interactions with the order parameter 
cause the fermi modes to become massive outside of isolated points.  Such 
Fermi points are stable under small perturbations of the system.  Thus, in 
$d=3$, we can have stable Fermi surfaces (as in the free fermion system) or 
Fermi points (as in ${}^3$He-A); however, Fermi {\it lines\/} appear to be 
(generically) unstable (see e.g.~\cite{volovik}).  We will now argue that 
this is the beginning of a pattern, explained as a consequence of Bott 
periodicity in K-theory.  

To see that K-theory controls the stability of Fermi surfaces, we study the 
exact propagator in Euclidean time (and in the obvious notation),
\be
G^{i\sigma}{}_{i'\sigma'}(\bk,\omega)=\langle\psi^{i\sigma}(0,0)\;
\psi^\dagger_{i'\sigma'}(\bk,\omega)\rangle.
\label{eeprop}
\ee
Gapless excitations correspond to a pole in $G$ along some submanifold 
$\Sigma$, of some dimension $d-p$, in the $(\bk,\omega)$-space.  We will 
analyze the stability of such poles under small perturbations in the theory, 
i.e., perturbations which do not change $G$ qualitatively far from $\Sigma$. 
(The analysis of large perturbations, corresponding to possible instabilities 
due to relevant deformations of the fixed point, is outside of the scope 
of this paper.)

It will be convenient to introduce a collective index $(i\sigma)\equiv a$, 
$a=1,\ldots N$ with $N=2^{[d/2]}\cdot n$, and consider the inverse exact 
propagator 
\bea
\CG_a{}^{a'}&\equiv&(G^{-1})_a{}^{a'}(\bk,\omega)\nonumber\\
&=&\delta_a^{a'}\left(i\omega-\bk^2/2m+\mu\right)+\Pi_a{}^{a'}(\bk,\omega),
\label{eepropcont}
\eea
with $\Pi_a{}^{a'}(\bk,\omega)$ the exact polarization tensor.  The question 
of stability of the manifold $\Sigma$ of gapless modes reduces to the 
classification of the zeros of the matrix $\CG$ that cannot be lifted by small 
perturbations of $\Pi_a{}^{a'}$.  Our arguments will be topological, and 
our results thus independent of the precise details of $\Pi_a{}^{a'}$.  

Assume that $\CG$ has a zero (i.e., $\det\CG$ vanishes) along a submanifold 
$\Sigma$ of dimension $d-p$ in the $d+1$-dimensional $(\bk,\omega)$ space.  
$\Sigma$ lies within the subspace of zero frequency, $\omega=0$.   Pick a 
point $\bk_F$ on the Fermi surface $\Sigma$, and consider the $p+1$ 
dimensions $\bk_\perp$ transverse to $\Sigma$ in the $(\bk,\omega)$-space at 
$\bk_F$.  A small perturbation of the system can either move the zero of 
$\CG$ slightly away from $\bk_F$ along $\bk_\perp$, or eliminate 
the zero altogether.  In the latter case, the purported Fermi surface is 
unstable, and a small perturbation will either produce a gap or will at least 
further reduce the dimension of the Fermi surface.   In order to classify 
{\it stable\/} zeros, consider a sphere $S^p$ wrapped around $\Sigma$ at some 
small distance $\epsilon$ in the transverse $p+1$ dimensions $\bk_\perp$, 
$|\bk_\perp-\bk_F|=\epsilon$.  We assume $\epsilon$ small enough so that 
this $S^p$ does not intersect any other components of the Fermi surface, a 
situation that can always be arranged in a generic point on the Fermi 
surface.  The matrix $\CG$ is nondegenerate along this $S^p$, and therefore 
defines a map 
\be \CG: S^p\rightarrow GL(N,\C)
\label{eemap}
\ee
from $S^p$ to the group of non-degenerate complex $N\times N$ matrices.   
If this map represents a nontrivial class in the $p$-th homotopy group 
$\pi_p(GL(N,\C))$, the zero along $\Sigma$ cannot be lifted by a small 
deformation of the theory.  The Fermi surface is then stable under small 
perturbations, and the corresponding nontrivial element of $\pi_p(GL(N,\C))$  
represents the topological invariant (or ``winding number'') responsible for 
the stability of the Fermi surface.  

One may worry that this could result in a very complicated pattern, 
dependent on the specific values of $p$ and $N$.  Fortunately, this is not the 
case, and the pattern that emerges is quite simple.  The key observation is 
that the values of $N$ and $p$ are always in the so-called stable regime 
\cite{karoubi}, of $N$ large enough so that $\pi_p(GL(N,\C))$ is independent 
of $N$.  This stable regime lies at $N>p/2$ \cite{karoubi}.  It is easy to 
check that in our setting, the stability condition is always satisfied. 

It is a deep mathematical result that in this stable regime, the homotopy 
groups of $GL(N,\C)$ define a generalized cohomology theory, known as 
K-theory \cite{karoubi}.  In K-theory, any smooth manifold $X$ is assigned an 
abelian group $\K(X)$.  (For $X$ noncompact, $\K(X)$ is to be interpreted as 
compact K-theory \cite{karoubi}.)  For example, for $X=\R^k$ this group is 
given by 
\be
\K(\R^{k})=\pi_{k-1}(GL(N,\C))
\label{eeksphere}
\ee
with $N$ in the stable regime.  The corresponding groups are known to be
$\K(\R^{2\ell})=\Z$ and $\K(\R^{2\ell+1})=0$.  This periodicity by two 
is known as Bott periodicity \cite{karoubi}.  

Our analysis of the Fermi surface can now be reinterpreted as a statement 
about K-theory: The map (\ref{eemap}) defines an element of the K-theory 
group $\K(\R^k)$ of the transverse space $\bk_\perp$; the Fermi surface 
$\Sigma$ is stable if this element is nontrivial.  We have established our 
first result:

{\it Stable Fermi surfaces in Fermi liquids are classified by K-theory; in the 
case of complex fermions, Fermi surfaces of codimension $p+1$ in the 
$(\bk,\omega)$-space are stable for $p$ odd, and unstable for $p$ even.}  
In $d=3$, this reproduces the observed pattern of stability mentioned above.  

As our first application of this picture, we will use K-theory to determine 
the dispersion relation of the gapless modes near a general stable Fermi 
surface $\Sigma$.  Such modes will be described by coarse-grained 
fermions $\chi^\alpha(\omega,\bp,\theta)$, with $\theta$ denoting coordinates 
on $\Sigma$, and $\bp$ being the spatial momenta normal to $\Sigma$.  The 
index $\alpha$ goes over some subset, to be determined below, of the range of 
$a$.  The leading quadratic part of the action is 
\be
S=\int d\mu(\omega,\bp,\theta)\left(\chi^\dagger_\alpha
\CD^\alpha{}_\beta(\omega,\bp,\theta)\chi^\beta+\ldots\right).
\label{eeleading}
\ee
$d\mu(\omega,\bp,\theta)$ is the flat measure $d\omega d^d\bk$ written 
in terms of the new coordinates $(\omega,\bp,\theta)$, and ``$\ldots$'' 
refers to interactions of $\chi^\alpha$, to be studied by RG methods in the 
vicinity of the free-field fixed point given by (\ref{eeleading}). $\CD$ 
is the operator we now wish to determine.    

K-theory provides an explicit construction of the generator $e$ in 
$\K(\R^{2\ell})$, known as the Atiyah-Bott-Shapiro (ABS) construction 
\cite{abs,karoubi,spingeo}.  Consider a stable Fermi surface $\Sigma$, 
of codimension $p+1\equiv 2\ell$, with winding number one.  
Any $\Sigma$ with a higher winding number $n$ can be perturbed into $n$ 
separated Fermi surfaces of winding number one.  The range $N$ of the index 
$a$ carried by the microscopic fermion is in the stable regime (and therefore 
quite large), but the range $\tilde N$ of the index $\alpha$ carried by 
the coarse grained fermions $\chi^\alpha$ can in principle be lower than $N$.  
The ABS construction determines the universal value of $\tilde N$ to be 
$\tilde N=2^{[p/2]}$.  If $p_\mu$ ($\mu=0,\ldots p$) are the dimensions 
transverse to $\Sigma$ in $(\bk,\omega)$, we first define the gamma matrices 
$\Gamma^\mu$ of $SO(p,1)$ to satisfy $\{\Gamma^\mu,\Gamma^\nu\}=
2\eta^{\mu\nu}$ with $\eta_{\mu\nu}$ a (Lorentz-signature) metric.  The ABS 
construction then gives the leading expression for the topological invariant 
$e$, and hence for the inverse propagator $\CD$ of the coarse-grained 
fermions $\chi^\alpha$ near the Fermi surface, 
\be
\CD=\Gamma^\mu p_\mu+\ldots.
\label{eegener}
\ee
The ``$\ldots$'' in (\ref{eegener}) refers to higher-order corrections to the 
leading low-energy term.  The precise form of the metric $\eta_{\mu\nu}$ is 
determined by the microphysics of $\psi$.  This establishes our second result: 

{\it At low energies, the dispersion relation of the coarse-grained gapless 
fermions $\chi^\alpha$ near the Fermi surface is governed by the 
Atiyah-Bott-hapiro construction of K-theory.}

The ABS construction has determined the unversal value of the range $\tilde N$ 
of $\alpha$, making $\chi^\alpha$ automatically a spinor of the Lorentz group 
$SO(p,1)$.  (Notice that $\tilde N\equiv 2^{[p/2]}\leq N$, with the equality 
only when $n=1$ and $p=d$.)  The equation of motion of $\chi^\alpha$ 
{\it in the low-energy regime} is the relativistic Dirac equation,
$\Gamma^\mu\p_\mu\chi=0+\ldots$, with ``$\ldots$'' now denoting possible 
nonrelativistic corrections at higher energies.  In the free theory, we 
get one copy of the Dirac fermion for each point (or patch) $\theta$ on the 
Fermi surface.  This is our third result: 

{\it The low-energy modes exhibit an emergent relativistic dispersion 
relation, in the $p+1$ dimensions transverse to the Fermi surface.}

Again, some low-dimensional examples of this behavior are well-known.  Here  
we have extended the statement to arbitrary dimensions, and found the 
emergent relativistic invariance as a simple consequence of the ABS 
construction.  Notice that both the gamma matrices {\it and\/} the fact that 
the coarse-grained fermions $\chi^\alpha$ transform as spinors of $SO(p,1)$ 
are emergent properties.  In particular, the spin-statistics theorem 
familiar from relativistic quantum field theory emerges naturally.  

The ABS construction played a crucial role in string theory, in Sen's picture 
\cite{sen} of stable D-branes as topological defects in the tachyon on 
higher-dimensional D-branes.  In the string theory literature, this 
construction is sometimes referred to as the ``$\Gamma\cdot x$ 
construction,''  since $x$ are the spacetime dimensions transverse 
to the soliton inside the higher-dimensional branes.  In the theory of 
Fermi liquids, we have put the $\Gamma\cdot x$ construction where it naturally 
belongs: in the momentum space.  

So far we have analyzed the theory locally near a point $\bk_F$ on the Fermi 
surface.  K-theory provides a natural extension of our arguments globally, 
to a Fermi surface of arbitrary topology.  Our construction demonstrates 
that a Fermi surface $\Sigma$ is stable precisely when it carries a 
topological charge in K-theory. Fermi surfaces are objects in K-theory and 
not just submanifolds in the $(\bk,\omega)$-space.  K-theory thus provides the 
natural arena for understanding the structure of Fermi surfaces.  We expect 
it to be particularly useful in the case of Fermi surfaces with complicated 
topologies and/or singularities. It could also be a natural tool for 
understanding the ideas of topological order in Fermi systems \cite{wenzee}.  

The case of real fermions is perhaps even more interesting.  Repeating the 
above steps for $\psi^{i\sigma}$ satisfying a reality condition 
$\psi^\ast\sim\psi$, we are naturally led to the classification of stable 
Fermi surfaces in terms of Real KR-theory \cite{karoubi}.  
The simplest reality condition in K-theory would define what is known as 
KO-theory, related to the homotopy groups of $GL(N,\R)$.  The subtlety here 
is that the involution that defines the reality condition on $\psi$ acts 
simultaneously as the time reversal symmetry, so as to preserve the equation 
of motion $i\p_t\psi+\Delta\psi/2m+\ldots=0$.  Consequently, the stable 
Fermi surfaces of dimension $d-p$ are now classified by groups 
$\KR(\R^{p,1})$.  
These groups are periodic in $p$ with periodicity 8; for low values of $p$, 
one finds $\Z$ for $p=1,5$, $\Z_2$ for $p=2,3$, and $0$ for $p=4,6,7,8$. 
The ABS construction of the low-energy dispersion relation is again given by 
(\ref{eegener}), and leads to coarse-grained relativistic Majorana fermions; 
the case of pseudo-Majorana fermions can be similarly incorporated.  

The novel phenomenon for real fermions is the existence of Fermi surfaces with 
a $\Z_2$ charge.  For example, in $2+1$ dimensions our framework predicts a 
stable Fermi line due to the $\Z$ charge in $\KR(\R^{1,1})$, but also a stable 
Fermi {\it point\/} carrying a $\Z_2$ charge in $\KR(\R^{2,1})$.  The 
low-energy dispersion relation is that of a relativistic $SO(2,1)$ Majorana 
fermion.  Since the charge takes values in $\Z_2$, two such Fermi points when 
brought together would annihilate, and a the system would form a gap.  Such 
behavior has been observed \cite{nread}.  Similarly, for real fermions in 
$d=3$ we can have a Fermi surface carrying a $\Z$ charge, but also a Fermi 
line and a Fermi point, each supported by a $\Z_2$ charge in KR-theory.  

Having analyzed the stability of the Fermi surface in the ground state, 
one can extend our framework to include the classification of topological 
defects in Fermi liquids.  In the semiclassical regime of small $\hbar$, one 
can consider slowly-varying Fermi surfaces (in the spirit of \cite{haldane}) 
and defects \cite{voletal} as submanifolds in $(\bx,t,\bk,\omega)$.  Repeating 
the analysis of this paper will now lead to a classification of the spectrum 
of stable topological defects, and the dispersion relations of their 
low-energy modes, again in terms of K-theory.  This pattern is indeed very 
reminiscent of how K-theory controls the spectrum of stable defects (in 
particular, D-branes) in string theory.  It remains to be seen whether this 
analogy between Fermi liquids and string theory runs deeper than suggested 
by the results of this paper.  

\acknowledgements

This paper is based on work largely done in Fall 2000 at Rutgers University.  
I wish to thank Tom Banks, Mike Douglas, Greg Moore, and Nick Read for 
valuable discussions at that time.   This work has been supported in part 
by NSF grant PHY-0244900, by the Berkeley Center for Theoretical Physics, 
and by DOE grant DE-AC03-76SF00098.  Any opinions, findings, and conclusions 
or recommendations expressed in this material are those of the author and 
do not necessarily reflect the views of the National Science Foundation.

\end{document}